\documentstyle[preprint,aps,prd]{revtex}

\begin{document}

\title {Coarse graining and decoherence in quantum field theory}
    
\author{Fernando Lombardo{\footnote{e-mail: lombardo@df.uba.ar}}}

\address{{\it
Departamento de F\'\i sica, Facultad de Ciencias Exactas y Naturales\\ 
Universidad de Buenos Aires- Ciudad Universitaria, Pabell\' on I\\ 
1428 Buenos Aires, Argentina}}

\author{Francisco D. Mazzitelli{\footnote{e-mail: fmazzi@df.uba.ar}}}

\address{{\it
Departamento de F\'\i sica, Facultad de Ciencias Exactas y Naturales\\ 
Universidad de Buenos Aires- Ciudad Universitaria, Pabell\' on I\\ 
1428 Buenos Aires, Argentina\\
and\\
Instituto de Astronom\'\i a y F\'\i sica del Espacio\\
Casilla de Correo 67 - Sucursal 28\\
1428 Buenos Aires, Argentina}}

\maketitle

\begin{abstract}
We consider a $\lambda \phi^4$ theory in Minkowski spacetime. 
We compute a ``coarse grained effective action" by integrating out the field 
modes with wavelength shorter than a critical value. 
From this effective action we obtain the evolution equation
for the reduced density matrix (master equation).
We compute the diffusion coefficients of this equation and
analyze the decoherence induced on the long-
wavelength modes. 
We generalize the results to the case of a conformally
coupled scalar field in DeSitter spacetime.
We show that the decoherence is effective as long as the 
critical wavelength is taken to be not shorter than the Hubble 
radius.  
\end{abstract}

\narrowtext
\newpage

\section{INTRODUCTION}

The quantum to classical transition \cite{review} is a very old and 
interesting
problem relevant in many branches of physics. It involves the concepts of 
{\it correlations}, i.e.
the Wigner function of the quantum system should have a peak
at the classical trajectories, and {\it decoherence}, that is, 
there 
should be no interference between classical trajectories. 
The density matrix should be diagonal.
In order to understand the emergence of classical behavior,
it is essential to consider the interaction 
of the system
with its environment \cite{unzu}, since
both the loss of quantum coherence and the onset of classical 
correlations depend strongly on this interaction \cite{laflou,pazsin}.

In the last years, there has been a renewed interest in the 
study of this transition in the context of quantum cosmology \cite{qc}.
The efforts are directed to explain the emergence of a classical
space-time metric from a full quantum theory, and to derive
the Einstein equations, or the quantum-corrected version of them.
The relevance of statistical  
and quantum effects such as noise, dissipation, particle creation,
back-reaction, etc,
have been subsequently elucidated 
in the context of quantum Brownian motion (QBM) \cite{hpz} and quantum
field theory \cite{ch,mori,GR}.

The quantum to classical transition is also of interest
in the standard cosmology. 
According to the inflationary theory,
quantum fluctuations of matter fields are the origin
of very small inhomogeneities which gave rise to the formation of
structures in the Universe \cite{varios}. A crucial assumption in this 
scenario is that quantum fluctuations become classical as
their wavelength become larger than the Hubble radius $H^{-1}$.
Therefore, a detailed analysis of this classical to quantum transition
is needed in order to complete our understanding of the
mechanism of structure-formation.

There has been several attempts in order to prove the previously 
mentioned
assumption. In Ref. \cite{blm}, this problem has been analyzed
using  a toy model which consists of two massless fields
interacting through a mixing of the kinetic terms. Similar theories
involving interactions with mass mixing terms have
been considered in Ref. \cite{felkam}. In all these models, the
interaction terms are quadratic and disappear after a redefinition
of the fields. In Ref. \cite{belga}, the same problem
has been studied in a theory consisting of two scalar fields with
a biquadratic interaction.

The  aim of this paper is to analyze the emergence of classical
inhomogeneities from quantum fluctuations in a more realistic model.
To begin with, we will consider a $\lambda \phi^4$ theory in 
Minkowski spacetime. 
Following Starobinsky's original suggestion for the stochastic 
inflationary model\cite{staro}, we will split
the  scalar field into two parts: the
 {\it system} $\phi_<$ containing the field modes whose
wavelengths are longer than certain fixed cutoff ($\Lambda^{-1}$), and the 
{\it environment} $\phi_>$ which contains the modes with wavelengths
shorter than the cutoff.  Although this splitting will cause
some technical difficulties, we will see that it is possible
to study and prove the assertion that quantum fluctuations
with  wavelengths longer than $\Lambda^{-1}$ decohere due to the 
interaction with
the environment. Moreover, it will be also possible 
to extend the results to curved spaces. For a conformally
coupled scalar field in a DeSitter spacetime, we will show that
Starobinsky's splitting between system and environment
is the most adequate in the following sense:
the minimum value for the critical wavelength
such that the system looses coherence is 
$\Lambda^{-1}\sim H^{-1}$. For shorter critical
wavelengths there are modes in the system that do not 
decohere.

The paper is organized as follows. In the next Section
we review the relation between the reduced density matrix for
the system and the Feynman-Vernon influence functional  
\cite {belga,feynver}. We integrate the environment 
degrees of freedom and
compute this functional perturbatively in $\lambda$,
for 
an arbitrary critical wavelength $\Lambda^{-1}$. 
In Section 3, we compute the propagator for the reduced
density matrix and indicate the procedure to obtain the master equation. 
In particular we compute the diffusion coefficients, 
since they give  the main contribution to the
decoherence.  Section 4 contains the
analysis of the loss of coherence produced by  each diffusion 
coefficient,
with particular emphasis on the cutoff-dependence of the results.
We also extend 
the results to the case of a conformally coupled scalar field in DeSitter 
spacetime, and discuss whether the choice $\Lambda\sim H$ is relevant or not. 
Section 5 contains our final remarks.  

\section{The influence Functional and the Density Matrix}
\label{sec:level2}
Let us consider a quantum scalar field in a Minkowski spacetime 
with a quartic 
self-interaction.  The classical action is given by
\begin{equation}S[\phi] = \int d^4x \{{1\over{2}} \partial_\nu \phi(x) 
\partial^\nu \phi(x) - {1\over{2}} m^2 \phi^2(x) - {1\over{4!}} \lambda 
\phi^4(x)\},\label{action}\end{equation} 
where $m$ is the bare mass of the scalar field and $\lambda$ is the bare 
coupling constant. Let us make a system-environment field splitting 
\begin{equation}\phi(x) = \phi_<(x) + 
\phi_>(x),\label{splitting}\end{equation} where we define the system by 
\begin{equation}\phi_<(\vec x, t) = \int_{\vert \vec k\vert < \Lambda} 
{d^3\vec k\over{(2 \pi)^3}} \phi(\vec k, t) \exp{i \vec k . \vec 
x},\label{sys}\end{equation} and the environment by 
\begin{equation}\phi_>(\vec x, t) = \int_{\vert \vec k\vert > \Lambda} 
{d^3\vec k\over{(2 \pi)^3}} \phi(\vec k, t) \exp{i \vec k . \vec 
x}.\label{env}\end{equation} The system-field contains the modes with 
wavelengths longer than the critical value $\Lambda^{-1}$, while the bath or 
environment-field contains wavelengths shorter than $\Lambda^{-1}$.      

After the splitting, the total action (\ref{action}) can be written as
\begin{equation}S[\phi] = S_0[\phi_<] + S_0[\phi_>] + S_{int}[\phi_<, 
\phi_>],\label{actions}\end{equation}
where $S_0$ denotes the free field action and the interaction term 
is given by
\begin{eqnarray}S_{int}[\phi_<, \phi_>] = - \int d^4x 
\{{\lambda\over{4!}}\phi_<^4(x) 
&& + {\lambda\over{4!}}\phi_>^4(x) + {\lambda\over{4}} \phi_<^2(x) \phi_>
^2(x) \nonumber \\ &&+ {\lambda\over{6}} \phi_<^3(x) 
\phi_>(x) + {\lambda\over{6}} \phi_<(x) 
\phi_>^3(x)\}.\label{inter}\end{eqnarray}

The total density matrix (for the system and bath fields) is defined
by 
\begin{equation} \rho[\phi_<,\phi_>,\phi_<',\phi_>',t]=\langle\phi_< 
\phi_>\vert {\hat\rho} \vert \phi_<' 
\phi_>'\rangle,\label{matrix}
\end{equation}
where $\vert \phi_<\rangle$ and $\vert \phi_>\rangle$ are the eigenstates of 
the field operators ${\hat\phi}_<$ and ${\hat\phi}_>$, respectively.
For simplicity, we will assume that 
the interaction is turned on
at the  
initial time $t_0$ and that, at this time, the system and the environment are 
not correlated (we ignore, for the moment, the physical consequences of such a 
choice). Therefore, the total density matrix can be written as the product of 
the density matrix operator for the system and for the bath 
\begin{equation}{\hat\rho}[t_0] = {\hat\rho}_{<}[t_0] 
{\hat\rho}_{>}[t_0].\label{sincorr}\end{equation} We will further assume that 
the initial state of the environment is the vacuum. 

We are interested in the influence of the environment on the evolution of the 
system. Therefore the reduced 
density matrix is the object of relevance. It is defined by 
\begin{equation}\rho_{red}[\phi_<,\phi'_<,t] = \int {\cal D}\phi_> 
\rho[\phi_<,\phi_>,\phi_<',\phi_>,t].\label{red}
\end{equation}
The reduced density matrix evolves in time by means of
\begin{equation}\rho_r[\phi_{<f},\phi_{<f}',t] = \int d\phi_{<i}\int 
d\phi_{<i}' J_r[\phi_{<f},\phi_{<f}',t\vert \phi_{<i},\phi_{<i}',t_0] 
\rho_r[\phi_{<i}\phi_{<i}',t_0],\label{evol}
\end{equation} 
where $J_r[t,t_0]$ is the reduced evolution operator 
\begin{equation}J_r[\phi_{<f},\phi_{<f}',t\vert \phi_{<i},\phi_{<i}',t_0] = 
\int_{\phi_{<i}}^{\phi_{<f}}{\cal D}\phi_< \int_{\phi'_{<i}}^{\phi_{<f}'}{\cal 
D}\phi_<' \exp{{i\over{\hbar}}\{S[\phi_<] - S[\phi_<']\}} 
F[\phi_<,\phi_<'].\label{evolred}
\end{equation}
The  influence functional (or Feynman-Vernon functional) 
$F[\phi_<,\phi_<']$ is defined as 
\begin{eqnarray}
F[\phi_<,\phi_<']= &&\int d\phi_{>i} \int 
d\phi_{>i}' \rho_{>}[\phi_{>i},\phi_{>i}',t_0] \int d\phi_{>f}
\int_{\phi_{>i}}^{\phi_{>f}}{\cal D}\phi_>\int_{\phi_{>i}'}^{\phi_{>f}}{\cal 
D}\phi_>'\nonumber \\
&&\times \exp {{i\over{\hbar}}\{S[\phi_>] + S_{int}[\phi_<,\phi_>] - 
S[\phi'_>] - S_{int}[\phi'_< , \phi'_>]} \}.
\end{eqnarray}
This functional takes into account the effect of the 
environment on the system. 

We 
define the influence action $\delta A[\phi_<,\phi_<']$ 
and the coarse grained effective action (CGEA) $A[\phi_<,\phi_<']$ as
\begin{equation}F[\phi_<,\phi_<'] = \exp {{i\over{\hbar}} 
\delta A[\phi_<,\phi_<']},\label{IA}\end{equation}
\begin{equation}A[\phi_<,\phi_<'] = S[\phi_<] - S[\phi_<'] + \delta 
A[\phi_<,\phi_<'].\label{CTPEA}\end{equation} 
We will calculate the influence action perturbatively in $\lambda$
and we will consider only terms up to order $\lambda^2$. The 
influence action has the following form
\begin{eqnarray}
\delta A[\phi_<,\phi_<'] = &&\{\langle 
S_{int}[\phi_<,\phi_>]\rangle_0 - \langle 
S_{int}[\phi_<',\phi_>']\rangle_0\}\nonumber \\
&&+{i\over{2}}\{\langle S_{int}^2[\phi_<,\phi_>]\rangle_0 - \big[\langle 
S_{int}[\phi_<,\phi_>]\rangle_0\big]^2\}\nonumber \\
&&- i\{\langle S_{int}[\phi_<,\phi_>] S_{int}[\phi_<',\phi_>']\rangle_0 - 
\langle S_{int}[\phi_<,\phi_>]\rangle_0\langle 
S_{int}[\phi_<',\phi_>']\rangle_0\} \label{inflac} \\
&&+{i\over{2}}\{S^2_{int}[\phi_<',\phi_>']\rangle_0 - 
\big[\langle 
S_{int}[\phi_<',\phi_>']\rangle_0\big]^2\},\nonumber
\end{eqnarray} 
where the quantum average of a functional of the fields is defined as
\begin{eqnarray}\langle B[\phi_>,\phi_>'] \rangle_0= \int 
d\phi_{>i}&& \int d\phi_{>i}' 
\rho_{>}[\phi_{>i},\phi'_{>i},t_0]\int d\phi_{>f}\nonumber \\ &&\times 
\int_{\phi_{>i}}^ {\phi_{>f}}{\cal D}\phi_>\int_{\phi_{>i}'}^{\phi_{>f}}{\cal 
D}\phi_>' \exp {{i\over{\hbar}}\{S_0[\phi_>] - S_0[\phi_>']\}} 
B.\label{averag}\end{eqnarray} 

We define the propagators of the environment field as
\begin{equation}\langle \phi_>(x),\phi_>(y)\rangle_0 = i 
G_{++}^\Lambda(x-y),\label{feyn}\end{equation}
\begin{equation}\langle \phi_>(x),\phi_>'(y)\rangle_0 = - i G_{+-}^\Lambda 
(x-y),\label{frecmas}\end{equation} 
\begin{equation}\langle 
\phi_>'(x),\phi_>'(y)\rangle_0 = - i G_{--
}^\Lambda(x-y).\label{dyson}\end{equation} 
These propagators are not the usual Feynman, positive-frequency Wightman, and 
Dyson propagators of the scalar field since, in this case, the momentum 
integration is restricted by the presence of the infrared cutoff $\Lambda$. 
The explicit expressions are  
\begin{equation}G_{++}^\Lambda (x-y)= \int_{\vert \vec p\vert > \Lambda} 
{d^4p\over{(2 \pi)^4}} e^{i p (x - y)} {1\over{p^2 - m^2 + i 
\epsilon}},\label{feypro}\end{equation} 
\begin{equation}G_{+-}^\Lambda (x-y) = \int_{\vert \vec p\vert > \Lambda} 
{d^4p\over{(2 \pi)^4}} e^{i p (x - y)} 2 \pi i 
\delta (x - y) \Theta(p^0),\label{frecmaspro}\end{equation}
\begin{equation}G_{--}^\Lambda (x-y) = \int_{\vert \vec p\vert > \Lambda} 
{d^4p\over{(2 \pi)^4}} e^{i p (x - y)} {1\over{p^2 - m^2 - i 
\epsilon}}.\label{dysonprop}\end{equation} 

As an example,  we show the 
expression for the propagator $G^\Lambda_{++}$
in the massless case. The usual Feynman propagator is
\begin{equation}G_{++}(x)={i\over{8 \pi^2}}{1\over{\sigma}} - 
{1\over{8 \pi}} 
\delta (\sigma),\nonumber \end{equation}
while 
\begin{eqnarray}
G^\Lambda_{++}(x)&& = {i\over{8 \pi^2}}\Big[{cos[\Lambda (r - t)]\over{r (r - 
t)}} + {cos[\Lambda (r + t)]\over{r(r+t)}}\Big] 
-{1\over{8 \pi^2}}\Big[{sin[\Lambda (r - t)]\over{r (r - t)}} - {sin[\Lambda 
(r + t)]\over{r (r + t)}}\Big]\nonumber \\ && \equiv G_{++}(x) - G_{++}^{\vert 
\vec p\vert <\Lambda}(x), \end{eqnarray} where $\sigma = {1\over{2}}x^2$ is 
one half of the geodesic distance. 

The influence action can be computed from Eqs. (\ref{inflac})-(\ref{dyson}) 
using standard techniques. After a long but straightforward calculation we 
find \begin{eqnarray} \delta A[\phi_<, \phi'_<]= && -\lambda \int d^4x 
\Big[{1\over{24}} (\phi^4_<(x) - \phi_<^{'4}(x)) + {1\over{4}} i 
G_{++}^\Lambda(0) (\phi_<^2(x) - \phi_<^{'2}(x))\Big]\nonumber \\ &&+ 
\lambda^2\int d^4x \int d^4y \Big[-{1\over{72}} \phi_<^3(x) G_{++}^ 
\Lambda(x-y) \phi_<^3(y) - {1\over{36}} \phi^3_<(x) G_{+-}^\Lambda(x-y) 
\phi_<^{'3}(y)\nonumber \\ && + {1\over{72}} \phi_<^{'3}(x) G_{--
}^\Lambda(x-y) \phi_<^{'3}(y) - {1\over{16}} \phi_<^2(x) i G_{++}^{\Lambda 
2}(x-y) \phi_<^2(y) \nonumber \\ && + {1\over{8}}\phi_<^2(x)i G_{+-}^{\Lambda 
2}(x-y) \phi_<^{'2}(y)-{1\over{16}}\phi_<^{'2}(x) i G_{--}^{\Lambda 2} 
(x-y)\phi_<^{'2}(y)\nonumber \\ && + {1\over{18}} \phi_<(x) G_{++}^{\Lambda 
3}(x-y)\phi_<(y) + {1\over{9}} \phi_<(x) G_{+-}^{\Lambda 
3}(x-y)\phi_<'(y)\nonumber \\ && - {1\over{18}}\phi_<'(x) G_{--}^{\Lambda 
3}(x-y) \phi'_<(y) \Big].\end{eqnarray} Defining $$P_\pm ={1\over{2}}(\phi^4_< 
\pm \phi'^4_<) ~~~;~~~ R_\pm ={1\over{2}}(\phi_<^3 \pm \phi'^3_<)$$ $$Q_\pm 
={1\over{2}}(\phi_<^2 \pm \phi'^2_<) ~~~;~~~ \Phi_\pm ={1\over{2}}(\phi_< \pm 
\phi'_<),$$ and using simple identities for the propagators, the 
real and imaginary parts of the 
influence 
action can be written as \begin{eqnarray} Re \delta 
A=&& - \lambda \int 
d^4x\{{1\over{12}}P_-(x) + {i\over{2}}G_{++}^\Lambda(0) Q_-(x)\}\nonumber \\ 
&&+ \lambda^2 \int d^4x\int d^4y \{-{1\over{18}} R_+(x) Re G^\Lambda_{++}(x-y) 
R_-(y) + {1\over{4}} Q_+(x) Im G^{\Lambda 2}_{++}(x-y) Q_-(y)\nonumber \\ &&+ 
{1\over{3}} \Phi_+(x) Re G^{\Lambda 3}_{++}(x-y) \Phi_-(y)\},\label{inff}\\
Im\delta A  =&&
\lambda^2 \int 
d^4x \int d^4y \{-{1\over{18}} R_-(x) Im G^\Lambda_{++}(x-y) R_-(y)\nonumber 
\\ &&-{1\over{4}} Q_-(x) Re G^{\Lambda 2}_{++}(x-y) Q_-(y) + {1\over{3}} 
\Phi_-(x) Im G^{\Lambda 3}_{++}(x-y) \Phi_-(y)\}, \end{eqnarray} 
this expression is valid in general but, for simplicity, in what follows we 
will consider only the massless case. 

The real part of the 
influence action in Eq.\ (\ref{inff}) contains divergences and must
be renormalized. As the propagators Eqs.\ (\ref{feyn})-(\ref{dyson})
differ from the usual ones only by the presence of the infrared
cutoff, the ultraviolet divergences coincide with those of
the usual $\lambda\phi^4$-theory. As a consequence, the 
influence action can be renormalized using the usual counterterms.

The term proportional to $G_{++}^{\Lambda}(0)Q_-(x)$ is divergent.
As $G_{++}^{\Lambda}= G_{++} - G_{++}^{\vert \vec p\vert <\Lambda}$
and $G_{++}^{\vert \vec p\vert <\Lambda}(0)$ is finite, the 
divergences of $G_{++}^{\Lambda}$ and  $G_{++}$ coincide.
This term renormalizes the mass.

Consider now the square of the Feynman propagator.
Using dimensional regularization we find
\begin{equation}G^{\Lambda 2}_{++}(x) = G_{++}^2(x) + G_{++}^{(\vert \vec 
p\vert <\Lambda) 2}(x) - 2 G_{++}(x) G_{++}^{(\vert \vec p\vert 
<\Lambda)}(x),\end{equation}
where
\begin{equation}G_{++}^2(x)={i\over{16 \pi^2}}[{1\over{n-4}}+\psi(1) - 4 
\pi]\delta^4(x)+i\Sigma(x) - \eta(x) - Log[4 \pi 
\mu^2],\nonumber\end{equation}
\begin{equation}\Sigma (x)={1\over{(2 \pi)^4}}\int d^4p e^{i p x} Log \vert 
p^2\vert,\nonumber\end{equation}
\begin{equation}\eta (x)={\pi\over{(2 \pi)^4}}\int d^4p e^{i p x} \Theta 
(p^2).\nonumber\end{equation}
Note that the divergence is the usual one, i.e., proportional
to ${1\over n-4}\delta^4(x-y)$ and independent of $\Lambda$.
Consequently, the term $Im G_{++}^{\Lambda 2}(x-y)Q_+(x)Q_-(y)$
in Eq.\ (\ref{inff}) is also divergent and renormalizes the coupling constant
$\lambda$. The other divergences can be treated in a similar 
way.
One can also check that the imaginary part of the effective
action does not contain divergences.

The real and imaginary parts of $\delta A[\phi_<,\phi_<']$ can be associated 
with the dissipation and  noise respectively, and can be
related by some integral equation known as the 
fluctuation-dissipation relation. 
                   
One can regard
the imaginary part of $\delta A$ as coming from three noise 
sources $\nu(x)$, $\xi(x)$, and $\eta(x)$ 
with a gaussian functional probability distribution given by 
\begin{eqnarray}
P[\nu(x), \xi(x), \eta(x)]=&& N_\nu N_\xi N_\eta \exp\bigg\{
-{1\over{2}}\int d^4x\int d^4y \nu(x)\Big[{\lambda^2\over{9}} 
Im G_{++}^\Lambda\Big]^{-1}\nu(y)\bigg\}\nonumber \\
&&\times \exp\bigg\{-{1\over{2}}\int d^4x\int d^4y \xi(x)
\Big[{\lambda^2\over{2}} Re G_{++}^{\Lambda 2}\Big]^{-1}\Big]
\xi(y)\bigg\}\nonumber \\
&&\times \exp\bigg\{-{1\over{2}}\eta(x)\Big[{-2\lambda^2\over{3}} 
Im G_{++}^{\Lambda 3}\Big]^{-1}\eta(y)\bigg\},\end{eqnarray}
where $N_\nu$, $N_\xi$, and $N_\eta$ are normalization factors. 
Indeed, we can write the imaginary part of the 
influence action as three functional integrals over the gaussian fields 
$\nu(x)$, $\xi(x)$, and $\eta(x)$ 
\begin{eqnarray}
\int {\cal D}\nu(x)\int {\cal D}&&\xi (x) \int {\cal D}\eta(x) 
P[\nu,\xi,\eta] \exp{ -{i\over{\hbar}} \bigg\{R_-(x)\nu(x) + Q_-(x) \xi(x) 
+ \Phi_-(x) \eta(x)}\bigg\}\nonumber \\
&&= \exp\bigg\{-{i\over{\hbar}}\int d^4x\int d^4y \
\Big[{\lambda^2\over{18}}R_-(x) ImG_{++}^\Lambda(x,y)R_-(y)\nonumber \\
&&+ {\lambda^2\over{4}}Q_-(x) Re G_{++}^{\Lambda 2}(x,y)Q_-(y) - 
{\lambda^2\over{3}}\Phi_-(x) Im G_{++}^{\Lambda 3}(x,y)\Phi_-(y)\
\Big]\bigg\}.\end{eqnarray}
Therefore, 
the CGEA can be rewritten as
\begin{equation}A[\phi_<,\phi_<']=-{1\over{i}} ln \int {\cal D}
\nu P[\nu]\int {\cal D}\xi P[\xi]\int {\cal D}\eta P[\eta] 
\exp\bigg\{i S_{eff}[\phi_<,\phi_<', \nu, \xi, \eta]\bigg\},\end{equation}
where 
\begin{equation}S_{eff}[\phi_<,\phi_<', \nu, \xi, \eta]= 
Re A[\phi_<,\phi_<']- \int d^4x\Big[R_-(x) \nu (x) + 
Q_-(x) \xi(x) + \Phi_-(x)\eta(x)\Big].\end{equation}
In the associated functional Langevin equation for the field \cite{mori,GR}, 
the corresponding stochastic force is
\begin{equation}F_{noise} \sim \nu(x) \phi_<^2(x) + \xi(x) 
\phi_<(x) + \eta(x),\end{equation}
i.e., multiplicative and additive noise. 
                 
To summarize this Section, we have seen that the environment 
has three fundamental 
effects over the system \cite{dowkerhall}:  renormalization, 
dissipation, and 
noise.

\section{The evolution of the reduced density matrix: master equation}
\label{sec:level3}

In this Section we will obtain the evolution equation for the
reduced density matrix (master equation). Before doing this,
let us briefly review the case of the QBM.
Denote by $x$ the coordinate
of the Brownian particle, by $\Omega$ its frequency,
and by $q_i$ the coordinates of the oscillators in the 
environment. For a linear coupling $x q_i$,
the  master equation for the reduced
density matrix $\rho_r(x,x',t)$ is of the form \cite{jppshort}
\begin{eqnarray}
i\hbar \rho_r(x,x',t)&=&\langle x\vert[H,\rho_r]\vert x'\rangle
-i\gamma(t)(\partial_x - \partial_{x'})\rho_r(x,x',t)\nonumber\\
&-&iD(t)(x-x')^2\rho_r(x,x',t)+f(t)(x-x')(\partial_x + \partial_{x'})
\rho_r(x,x',t),
\label{meqbm}
\end{eqnarray}
where the coefficients $\gamma (t), D(t)$ and $f(t)$ depend 
on the properties of the environment (temperature $\beta^{-1}$
and spectral density $I(\omega)$).
The difussion coefficient $D(t)$ 
is 
\begin{equation}
D(t)=\int_0^t ds~~cos (\Omega s)~~\int_0^{\infty}d\omega
~~I(\omega)coth({1\over 2}\beta\hbar\omega)~~cos(\omega s),
\label{dt}
\end{equation}
and gives the main contribution
to the decoherence. Indeed, an approximate solution
of Eq.\ (\ref{meqbm}) is
\begin{equation}
\rho_r[x,x';t] \approx \rho_r[x,x',0] \exp{\Big[-(x - 
x')^2\int_0^t D(s) ds}\Big],\label{qbmdecay}\end{equation}
and we see that the off-diagonal terms of the density matrix
are suppressed as long as $\int_0^t D(s) ds$ is large enough.

For non-linear couplings like $x^nq_i^m$, one expects 
the master equation to contain  
terms of the form
$iD^{(n,m)}(t)(x^n-x'^n)^2\rho_r$.

Here we will generalize these results to the field theory
of Section 2.
To this end, we will compute the 
reduced evolution operator Eq.\ (\ref{evolred}). As the operator is defined 
through a path integral, we can obtain an estimation using the saddle point 
approximation: 
\begin{equation}
J_r[\phi_{<f},\phi'_{<f},t\vert\phi_{<i},\phi'_{<i},t_0] 
\approx \exp{ {i\over{\hbar}} A[\phi_<^{cl},\phi_<^{'cl}]},
\label{prosadle}\end{equation}
where $\phi_<^{cl} (\phi_<^{' cl})$ is the solution of the equation of motion
${\delta Re A\over\delta\phi_<}\vert_{\phi_<=\phi'_<}=0$
with boundary conditions $\phi_<^{cl}(t_0)=\phi_{<i}(\phi'_{<i})$
and $\phi_<^{cl}(t)=\phi_{<f} (\phi'_{<f})$.

The decoherence effects are contained in the imaginary part 
of the CGEA, which is already of order $\lambda^2$.
As a consequence, in the evaluation of $Im A$ we can approximate $\phi_<^{cl}$ 
by the solution of the free field equation satisfying the appropiate boundary 
conditions. Therefore, the classical solution reads 
\begin{equation}\phi_<^{cl}(\vec x, s) =  [\phi_{<f} {sin(k_0 s)\over {sin(k_0 
t)}} + \phi_{<i} {sin[k_0 (t - s)]\over{sin(k_0 t)}}] cos(\vec k_0 . \vec 
x)\equiv \phi_<^{cl}(s) cos(\vec k_0 . \vec 
x)
,\label{classicphi}\end{equation} where we assumed that the system-field 
contains only one Fourier mode with $\vec k = \vec k_0$. This is a sort of 
``minisuperspace" approximation for the system-field that will greatly 
simplify the calculations.

Following the same techniques used for 
the QBM \cite{jppshort},
to obtain the master equation we must compute the time derivative of the 
propagator $J_r$, and eliminate the dependence on the initial field 
configurations $\phi_{<i}$ and $\phi'_{<i}$ that enters through 
$\phi_{<}^{cl}$ and $\phi_<^{'cl}$. 
This can be easily done because the free propagator, defined as 
\begin{equation}J_0[\phi_{<f}, \phi'_{<f}, t\vert \phi_{<i}, \phi'_{<i}, 0] = 
\int_{\phi_{<i}}^{\phi_{<f}}{\cal D}\phi_< \int_{\phi'_{<i}}^{\phi'_{<f}} 
{\cal D}\phi'_< \exp\{{i\over{\hbar}} [ S_0(\phi_<) - 
S_0(\phi'_<)]\},\label{propdeJ0}\end{equation}
satisfies the identities
\begin{equation}\phi_<^{cl}(\vec x, s) J_0 = \Big[ \phi_{<f} cos[k_0(t - s)] + 
{sin[k_0(t - s)]\over{k_0}} i 
\partial_{\phi_{<f}}\Big]J_0,\label{rel1}\end{equation}
and
\begin{equation}\phi_<^{'cl}(\vec x, s) J_0 = \Big[ \phi'_{<f} cos[k_0(t - s)] 
- {sin[k_0(t - s)]\over{k_0}} i 
\partial_{\phi'_{<f}}\Big]J_0.\label{rel2}\end{equation}

The temporal derivative is given by 
\begin{eqnarray}i\hbar \partial_t J_r[&&\phi_{<f},\phi'_{<f},t\vert 
\phi_{<i},\phi'_{<i},0] = \bigg\{h_{ren}[\phi_<] - h_{ren}[\phi'_<] - i 
\lambda^2 [{(\phi_{<f}^3 - \phi_{<f}^{'3})V\over{1152}}\nonumber \\ &&\times 
\int_0^t ds R_-^{cl}(s) Im G_{++}^{\Lambda}
(3k_0;t-s)\nonumber \\ && 
+ {(\phi_{<f}^2 - \phi_{<f}^{'2})V\over{32}} \int_0^t ds Q_-^{cl}
(s)(Re G_{++}^{\Lambda 2}(2k_0;t-s)+ 2 Re G_{++}^{\Lambda 2}(0;t-s))
\nonumber \\ && - {(\phi_{<f} 
- \phi'_{<f})V\over{6}} \int_0^t ds \Phi_-^{cl}(s) 
Im G_{++}^{\Lambda 3}(k_0;t-s)]\nonumber \\
&&- \lambda^2 [-{(\phi_{<f}^3 + \phi_{<f}^{'3})V\over{1152}}
\int_0^t ds R_-^{cl}(s) 
Re G_{++}^{\Lambda}(3k_0;t-s) \nonumber \\ && 
+ {(\phi_{<f}^2 + \phi_{<f}^{'2})V\over{32}} \int_0^t ds Q_-^{cl}
(s)(Im G_{++}^{\Lambda 2}(2k_0;t-s)+ 2 Im G_{++}^{\Lambda 2}
(0;t-s))\nonumber \\ && + {(\phi_{<f} 
- \phi'_{<f})V\over{6}} \int_0^t ds \Phi_-^{cl}(s) 
Re G_{++}^{\Lambda 3}(k_0;t-s)]+ ... \bigg\}J_r[\phi_{<f},\phi'_{<f},t\vert 
\phi_{<i},\phi'_{<i},0],\label{timeder} \end{eqnarray}  
where $G_{++}^{\Lambda n}(k;t-s)$ is the Fourier transform of 
the $n$-power of the propagator
($n =$ 1, 2, or 3). The ellipsis denotes other terms coming from the time 
derivative  that do not contribute to the diffusive effects (we will ignore 
all terms not proportional to $(\phi_{<f}^n - \phi_{<f}^{'n})^2$
in the final equation). The volume 
factor $V$ that appears in Eq. (\ref{timeder}) is due to our assumption that 
the system-field contains only one mode. As usual in quantum field theory, it 
disappears when considering properly defined wave packets.

Using the identities Eqs.\ (\ref{rel1}) and (\ref{rel2}) 
we can remove the dependence on the initial boundary conditions and
obtain the master equation.   
As the final equation is complicated and not very illuminating,
we only show the correction to the usual unitary evolution term 
coming from the noise kernels: 
\begin{eqnarray}i \hbar \partial_t \rho_r && [\phi_{<f},\phi'_{<f},t] = 
\langle \phi_{<f}\vert\Big[{\hat H}_{ren}, 
{\hat\rho}_r\Big]\vert\phi'_{<f}\rangle 
- i \lambda^2 [{(\phi_{<f}^3 
- \phi_{<f}^{'3})^2 V\over{1152}}D_1(k_0;t)\nonumber \\
&& + {(\phi_{<f}^2 - 
\phi_{<f}^{'2})^2 V\over{32}}D_2(k_0;t) - {(\phi_{<f} - 
\phi'_{<f})^2 V\over{6}}D_3(k_0;t)] \rho_r[\phi_{<f},\phi'_{<f},t] + ...~~ 
.\label{master}\end{eqnarray} This equation contains three time-dependent 
diffusion coefficients $D_i(t)$. Up to one loop, only $D_1$ and $D_2$ survive 
and are given by \begin{eqnarray}D_1(k_0;t)= && \int_0^t ds ~~ cos^3(k_0 s)Im 
G_{++}^{\Lambda}(3k_0;t-s)\nonumber \\ && = {1\over{6 k_0}}\int_0^t ds ~~ 
cos^3(k_0 s) ~~ cos(3 k_0 s) ~~ \theta(3 k_0 - \Lambda)\nonumber \\ && = {2 
k_0 t + 3 sin(2 k_0 t) + {3\over{2}} sin(4 k_0 t) + {1\over{3}} sin(6 k_0 
t)\over{576 k_0^2}},~~~~ {\Lambda\over{3}}<k_0<\Lambda\label{d1}\end{eqnarray} 

\begin{equation}D_2(k_0;t) = \int_0^t ds ~~ cos^2(k_0 s)(Re G_{++}^
{\Lambda 2}(2k_0;t-s)+ 
2Re G_{++}^{\Lambda 2}(0;t-s)).\label{d2prev}\end{equation} 
Using that
\begin{eqnarray}Re G_{++}^{\Lambda 2}(2k_0;t-s)= 
{\pi\over{k_0}}\bigg\{&& \int_\Lambda^{2k_0+\Lambda}dp\int_\Lambda^{2k_0 + p}dz 
cos[(p + z)s]\nonumber \\
&& + \int_{2k_0+\Lambda}^{\infty}dp \int_{p-2k_0}^{p+2k_0}dz 
cos[(p+z)s]\bigg\},\end{eqnarray} 
\begin{equation}Re G_{++}^{\Lambda 2}(0;t-s)= 
\pi\bigg\{2\pi \delta(s) - 2 {sin(2 \Lambda 
s)\over{s}}\bigg\},\end{equation} 
the $D_2$ diffusion coefficient reads 
\begin{eqnarray}D_2(k_0;t) = {\pi\over{4}}\bigg\{&&3 \pi - ({3\over{2}} - 
{\Lambda\over{2k_0}})Si[2 t(\Lambda - k_0)]\nonumber \\ && - (2 - 
{\Lambda\over{2 k_0}}) Si[2 \Lambda t] - ({3\over{2}} + 
{\Lambda\over{2k_0}})Si[2 t(\Lambda + k_0)] -(1 +
{\Lambda\over{2k_0}})Si[2t(2k_0 + \Lambda)]\nonumber \\ &&+ {cos[2\Lambda 
t]\over{4k_0t}} - {cos[2 t(\Lambda + k_0)]\over{4k_0t}}+{cos[2 t(\Lambda - 
k_0)]\over{4k_0t}}-{cos[2 
t(2k_0+\Lambda)]\over{4k_0t}},\label{d2}\end{eqnarray} where $Si[z]$ 
denotes 
the sine-integral function \cite{abramo}.

Eq.\ (\ref{master}) is the generalization of the QBM master 
equation we were looking for.
In our case, the system is coupled in a nonlinear form. Therefore, owing 
to the existence of three interaction terms ($\phi_<^3 \phi_>$, 
$\phi_<^2 \phi_>^2$, and $\phi_< \phi_>^3$) there are three diffusion 
coefficients in the master equation. The form of the 
coefficients is fixed by these couplings and by the particular
choice of the quantum state
of the environment.

Our results are valid in the single-mode approximation of 
Eq.\ (\ref{classicphi}). In this approximation one obtains a reduced density 
matrix for each mode $\vec{k}_0 $, and neglects the interaction 
between different system-modes. Due to this interaction, in the general case, 
$\rho_r$ will be different from $\prod_{\vec k_0} \rho_r(\vec k_0)$. This 
point deserves further study.


\section{Decoherence and Cosmological Perturbations} 
\label{sec:level4} 

In this Section we will analyze the behavior of the 
diffusion  coefficients $D_1$ and $D_2$.  We will 
also extend the results to DeSitter space, in order
to discuss the issue of the quantum to classical
transition of the primordial quantum fluctuations.

A detailed
analysis of the quantum to classical transition in the
model we are considering is a very complicated task.
One should 
analyze in detail the master equation and 
see whether the off-diagonal elements of the reduced density 
matrix are suppressed or not. One should also
study the form of the Wigner function, and see
whether it predicts classical correlations or not. This is 
beyond the scope of this paper.  Having in mind the analogy
with the QBM, here we will only be concerned 
with the diffusive terms of the master equation.
We will use the value of the diffusion coefficients
as the signal for decoherence, which is, of course,
only a rough approximation.

After these words of caution, let us turn to
the analysis of Eqns. (\ref{d1}) and (\ref{d2}).
The coefficient $D_1$ is associated to the interaction term
$\phi_<^3\phi_>$. We are considering only one Fourier mode
for the system, with wavevector $\vec k_0$. The
environment field enters linearly in the interaction term,
and contains only modes with $k>\Lambda$ (by definition).
Consequently, $\phi_<$ is only coupled with 
the $\vec k =3 \vec k_0$
mode of the environment. This implies that
$D_1$ is different from zero only if 
${\Lambda\over 3}<k_0<\Lambda$.  In Fig. 1 we plot
the temporal evolution of $D_1(k_0,t)$. The 
diffusive effects increase with time as ${t\over k_0}$
(of course our perturbative calculations are not valid
at large times). The shape of the curve depends on
$k_0$ and takes its maximum value for
$k_0={\Lambda\over 3}$. To illustrate this, in Fig. 2
we plot $D_1(k_0,t)$ as a function of
$k_0$ for a fixed time.

It is interesting to note that 
we would obtain a similar diffussion
coefficient in a particular QBM,
with interaction $x^3 q$ and spectral density
$I(\omega )\sim \delta (\omega - 3k_0) \theta (\omega -\Lambda)$.

Let us now consider the coefficient $D_2$ coming
from the interaction term $\phi_<^2\phi_>^2$.
As the interaction is now quadratic in $\phi_>$, there
are no restrictions on the values of $k_0$ such that
$D_2\neq 0$. In Fig. 3 we have plotted the temporal evolution
of this coefficient for several values of
$y={k_0\over\Lambda}$ (the variable $y$ labels the different
curves). Although formally the diffusion coefficients
should vanish at $t=0$ (see Eq. (\ref{d2prev})), the graph shows
an initial jolt for all values of $y$. The origin of this
behavior is the uncorrelated initial condition Eq.(\ref{sincorr}).
We have assumed that the interaction between the 
system and the environment is turned on instantaneously
at $t=0$. ``Instantaneously" means in a time scale shorter
than all time scales present in the system and 
environment. Therefore, this initial condition
assumes implicitly the existence of an ultraviolet cutoff
for the frequencies in the environment. Had we considered
such cutoff $\Lambda_{uv}$, $D_2$ would have vanished
at $t=0$ and developed a peak at $t\sim{1\over\Lambda_{uv}}$.
This is exactly what happens in the QBM \cite{hpz}. In order
to avoid this cutoff, one should consider a situation
where the interaction is turned on adiabatically.

Fig. 3 shows that, after the (probably unphysical) jolt,
the diffusive effects grow with $y$, and are maximum
for $k_0\simeq \Lambda$. The physical interpretation is that
the decoherence increases when the frequency
of the system is large enough to excite the environment.

Figs. 4 and 5 show the dependence of $D_2$ with $k_0$,
at a fixed time. For large values of $ l = \Lambda t$
$(\Lambda\rightarrow\infty)$, the environment contains only
very high frequencies and the diffusion coefficient is
very small unless $k_0 \sim \Lambda$ (Fig. 4). When $l \leq 1$,
$D_2$ is appreciably different from zero for all modes
inside the system, i.e., the environment is able to
produce decoherence for $0<k_0<\Lambda$ (Fig. 5).
Finally, for $l \ll 1$ $D_2(k_0,t)$ approaches a
non vanishing value in the limit $k_0<\Lambda\rightarrow 0$.

It is instructive to compare our results with
those of Ref. \cite{belga}, where a similar analysis
is done for two independent fields $\phi$ and
$\varphi$ coupled through a biquadratic
interaction $\varphi^2\phi^2$. In that case,
one can show that the diffusive effects are more
important in the infrared. In our case, the analysis
of the coefficients $D_1$ and $D_2$ shows that
the loss of coherence is more important
for $k_0\sim\Lambda$ rather than in the infrared. 

After this long journey, we now turn to the issue
of cosmological perturbations. The results of 
Section 3 can be easily generalized to the case of
a conformally coupled scalar field  in a Robertson Walker
spacetime. The expressions for the diffusion coefficients
$D_1$ and $D_2$ are the same as those for Minkowski space
Eqns. (\ref{d1}) and (\ref{d2}),
with the replacement $t\rightarrow \eta=\int_0^t {dt'\over a(t')}$.
Here $a(t)$ is the scale factor of the Robertson Walker
metric and $\eta$ is the conformal time. For the particular
case of a DeSitter spacetime we have $a(t)=\exp (Ht)$ and 
$\eta={1\over H }(1-{1\over a})$. Therefore, the dimensionless
quantities $\Lambda t$ and $k_0 t$ in Eqs. (\ref{d1}) and (\ref{d2})
become ${\Lambda\over H }(1 - {1\over{a}})$
and
${k_0\over H}(1 - {1\over{a}})$ respectively. As $a(t=0)=1$, $\Lambda$ and $k_0$ are the 
physical values of the cutoff and wavevector at the initial time.

Now we can ask the following question: which is the maximum
value of the cutoff $\Lambda$ such that, a few e-foldings after the
initial time, all modes with $k_0\leq\Lambda$ still suffer the diffusive
effects? The analysis of the diffusion coefficients in Minkowski 
spacetime shows that $D_1$ is very small for $l \gg 1$ (Fig. 2). A similar
thing happens for $D_2$: if $l \gg 1$ this coefficient is very small in 
the infrared sector, and the diffusive effects are relevant only for 
$k_0 \sim \Lambda$ (Fig. 4). As in DeSitter spacetime we have $l \approx 
{\Lambda\over{H}}$, we conclude that we can include in our 
system only those modes with wavelength larger than 
the horizon $H^{-1}$. This is Starobinsky's original suggestion. If we 
include wavelengths shorter than the horizon, the low 
frequency sector of the system 
cannot excite the environment and does not decohere. 

Our results do not apply directly to the
inflationary models, since our scalar field is 
conformally coupled to the curvature. A more
realistic model should contain a minimally coupled 
scalar field. Moreover, the cutoff should be time
dependent \cite{habib} since the system should contain, at
each time, the modes with $k_{0ph}={k_0\over{a}}<H$. In spite of this, we 
think that this example illustrates
the main aspects of the problem.

\section{FINAL REMARKS}

Let us summarize the new results contained in this paper.
After the integration of the high frequency modes in Section 2,  
we obtained the CGEA for the 
low energy modes. From the imaginary part of the CGEA
we obtained, in Section 3, the diffusion coefficients
of the master equation.
Similar calculations have been done previously \cite{belga}
considering two scalar fields as system and environment.
Here the system and environment are two sectors of a single
scalar field, and the results depend on the  ``size" of
these sectors, which is fixed by
the critical wavelength $\Lambda^{-1}$. 

In Section 4 we analyzed
the $\Lambda$-dependence of the diffusion coefficients.
We showed that the decoherence is larger for those modes
in the system whose wavelength is close to 
the critical value.
We also generalized the results to DeSitter spacetime
and showed that, when the critical wavelength 
is equal to the horizon size, all modes in the
system  suffer decoherence and may become classical.

To end this paper, we would like to point
out an interesting connection between our work, the renormalization
group and the block-spin transformation. 
The CGEA  of Section 2 is similar to the
``average effective 
action" (AEA)  proposed in Ref.
\cite{wett} (see also Refs \cite{morris,liao}). The AEA is 
an effective action for averages of the field over a finite
space-time volume in Euclidean space. It is
defined through a splitting similar to the one used in this paper,
but now with a critical {\it Euclidean} four-momentum 
$p_{\mu}p_{\mu}=\Lambda^{2}$,
where $\Lambda^{-1}$ is the typical size of the Euclidean average volume. The 
AEA interpolates between the classical action ($\Lambda\rightarrow 0$) and the 
usual effective action ($\Lambda\rightarrow \infty$), and realizes the block-
spin transformation \cite{wilson} in the continuum limit. 

It is possible to write an exact evolution equation
for the dependence of the AEA with the scale $\Lambda$. This 
equation was originally discussed by Wegner and Houghton \cite{wh}, and by 
Polchinsky \cite{pol} in his proof of renormalizability of $\lambda \phi^4$-
theory. It can be used as the starting point for non perturbative calculations 
in quantum field theory. 

The CGEA we found in this paper can be seen as 
the Close Time Path version (as opposed to Euclidean version)
of the AEA, where the field averages are taken over
3-volumes contained in $t=const$ surfaces. Here we evaluated the CGEA
perturbatively. However, it is possible to
write an exact equation for the dependence of the CGEA
on the critical  wavelength. This evolution equation could be
the starting point for non perturbative evaluations
of the influence functional. Work in this direction is in
progress \cite{nosotros}.

\acknowledgments
We are greatly indebted to J.P. Paz for many useful
discussions. We have also benefited from
conversations with
E. Calzetta, B. L. Hu, L. Oxman, R. 
Scoccimarro, and M. Zaldarriaga. This research was 
supported by Universidad de Buenos Aires, Consejo Nacional de Investigaciones 
Cient\'\i ficas y T\'ecnicas and by Fundaci\' on Antorchas.  

\newpage

\hskip5cm FIGURE CAPTIONS

\bigskip

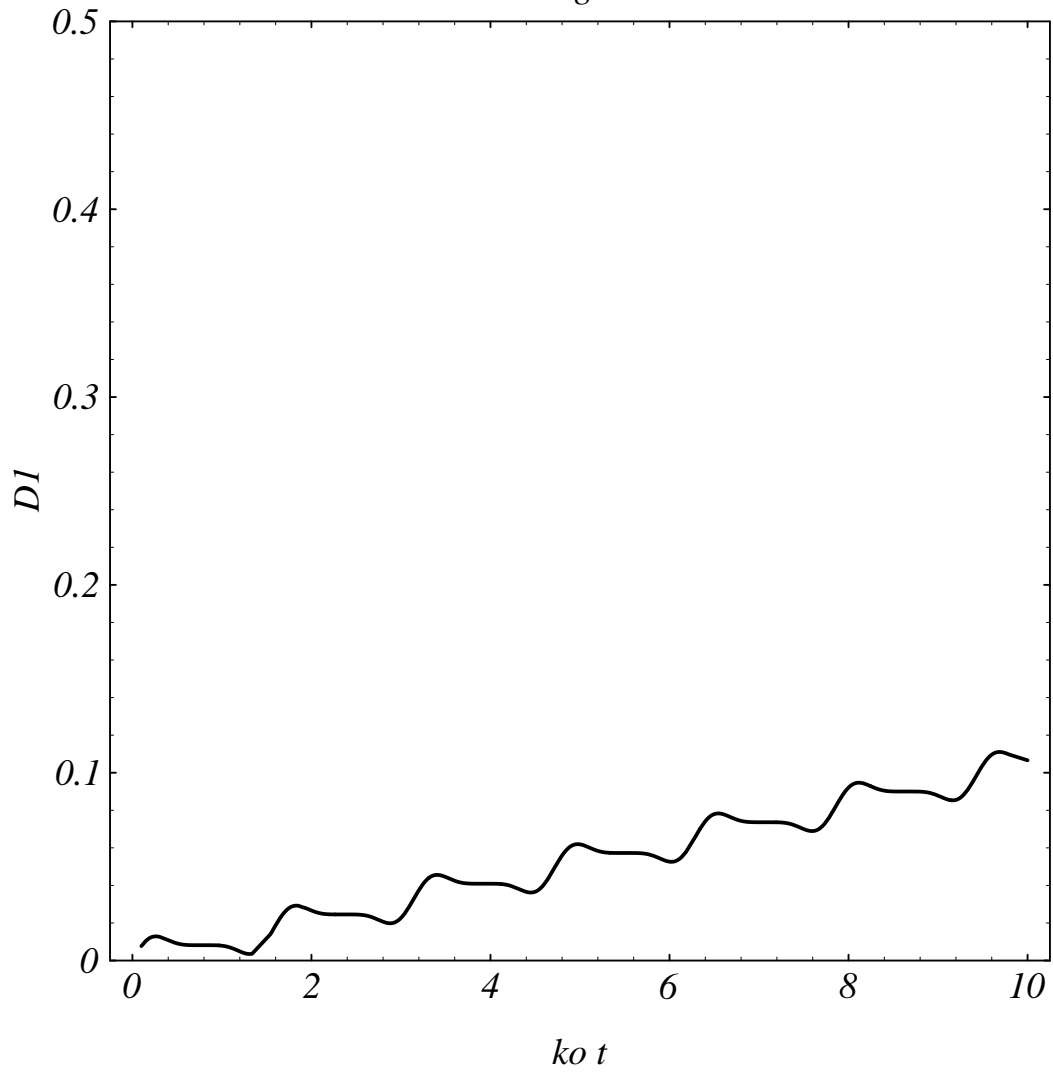
FIG.1. Temporal evolution of the coefficient $D_1$ for a fixed 
${\Lambda\over{3}} < k_0 < \Lambda$ 

\bigskip

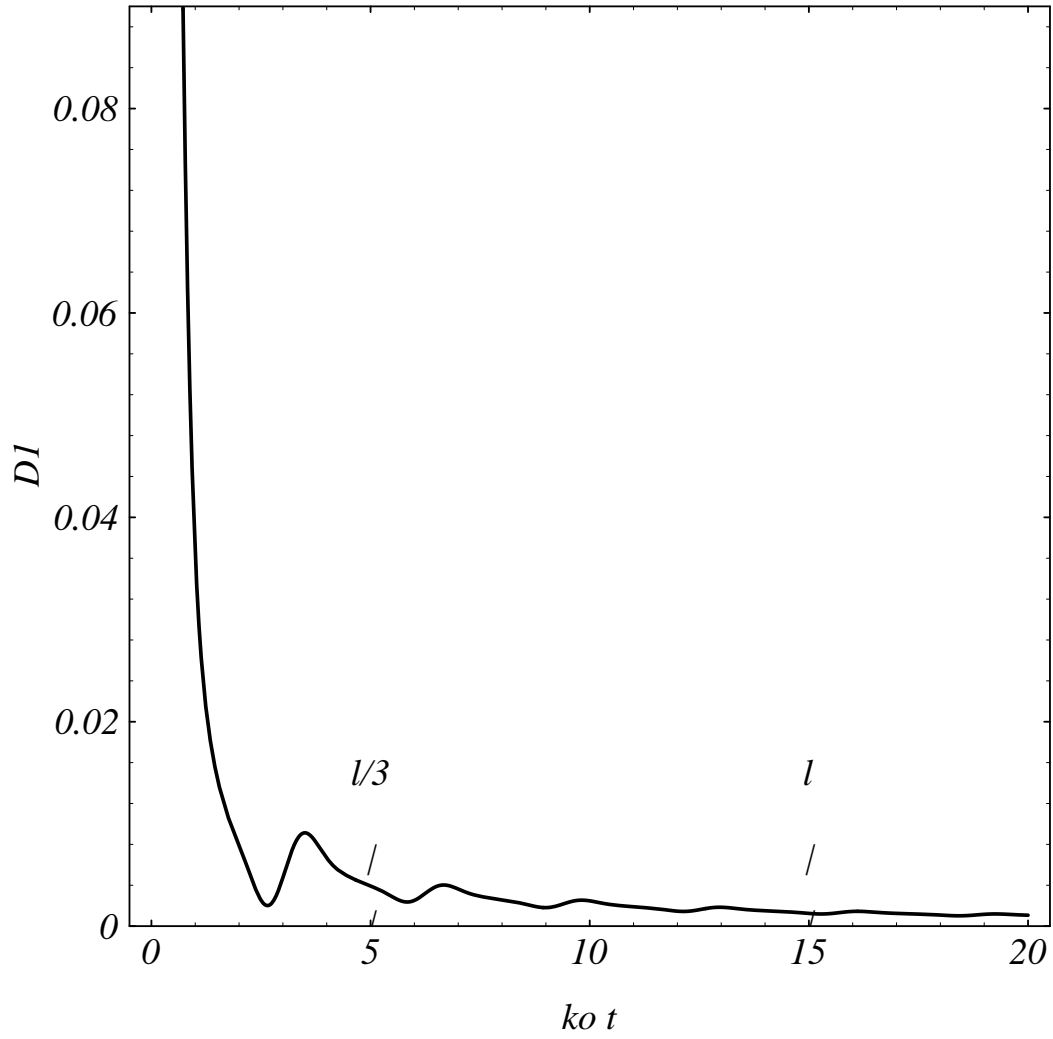
FIG.2. The coefficient $D_1$ as a function of $k_0$ for a fixed value of 
$t$. $D_1$ is different from zero only inside the window ${\Lambda\over{3}} < 
k_0 < \Lambda$. In the figure we indicated this window for the  particular 
value $l = \Lambda t = 15$

\bigskip

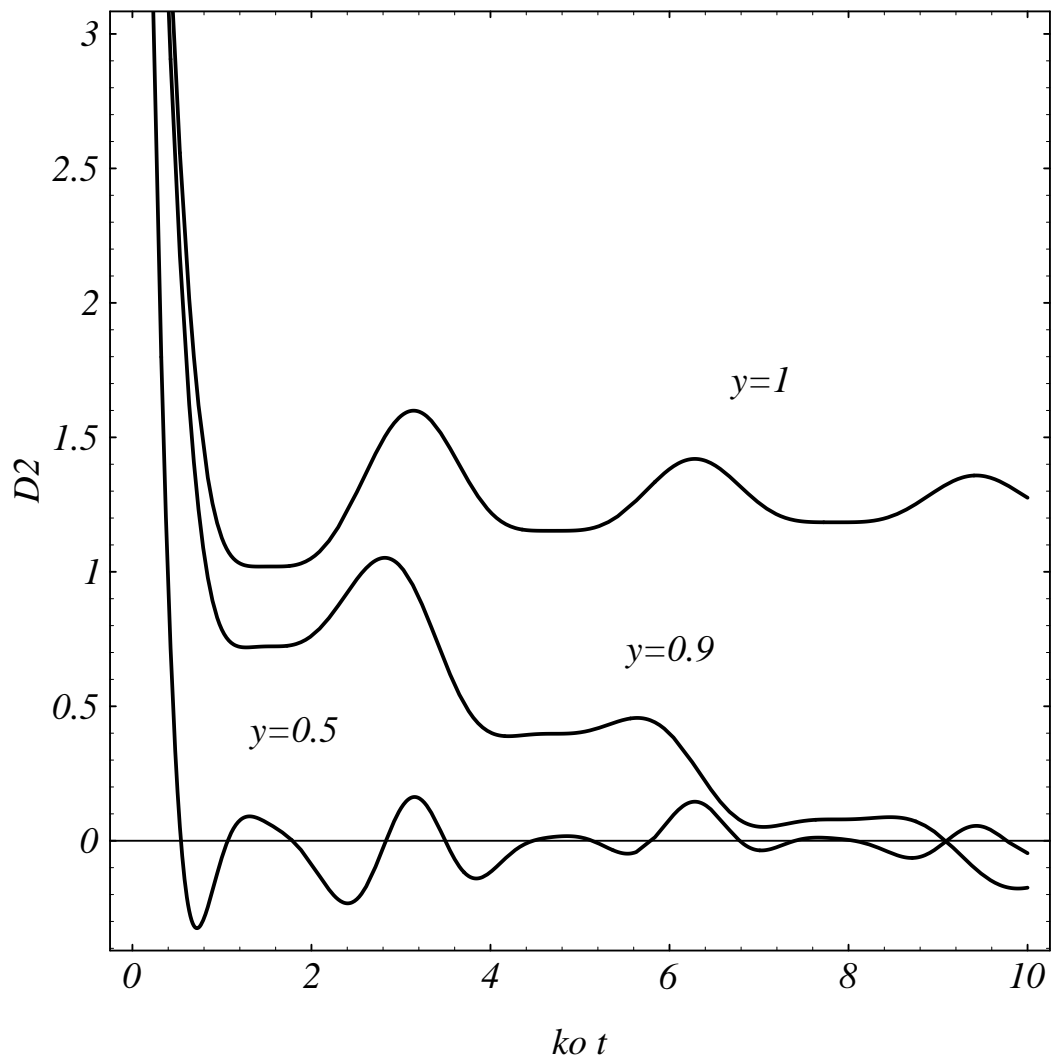
FIG.3. Temporal evolution of the coefficient $D_2$ for different values of 
$y = {k_0\over{\Lambda}}$ 

\bigskip

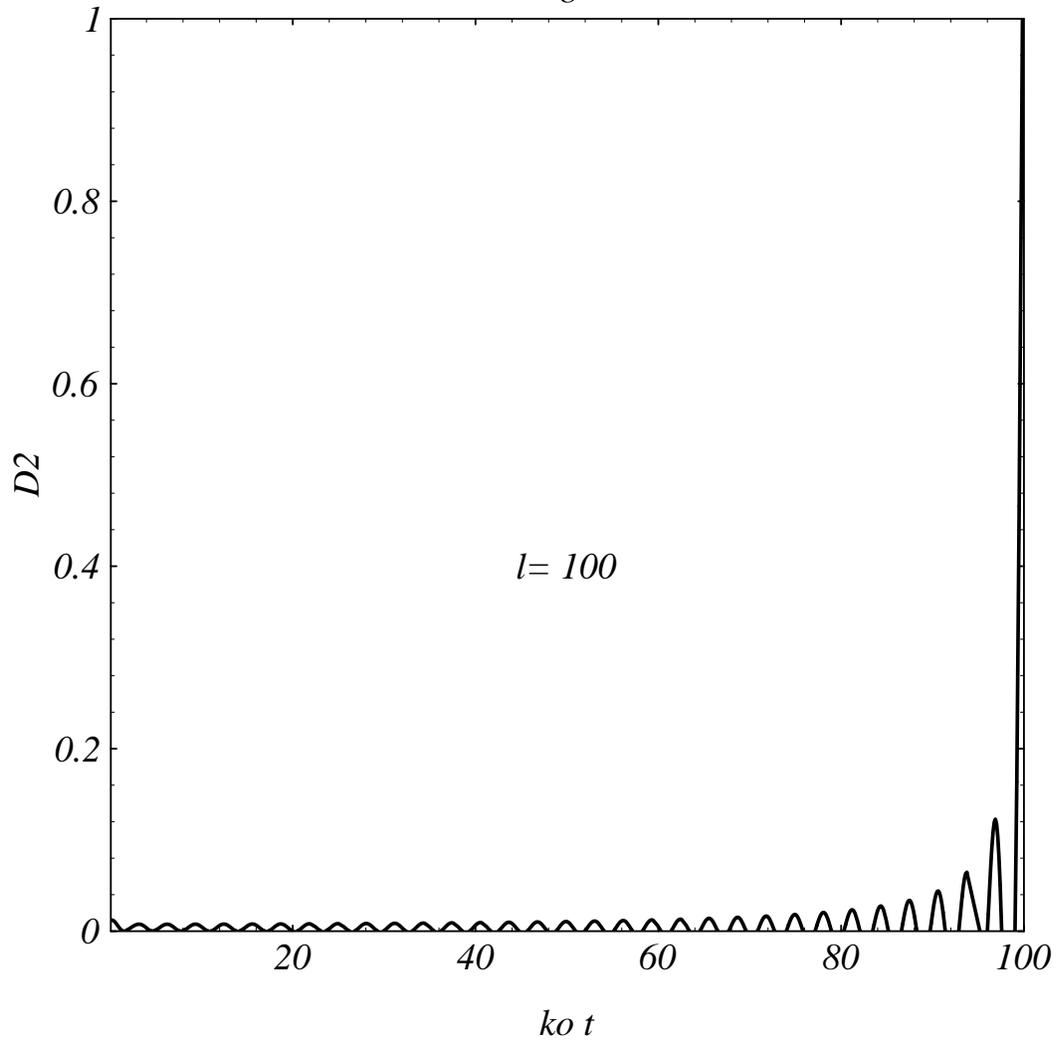
FIG.4. The coefficient $D_2$ as a function of $k_0$ for a fixed value of $t$ 
and $l = \Lambda t = 100$

\bigskip

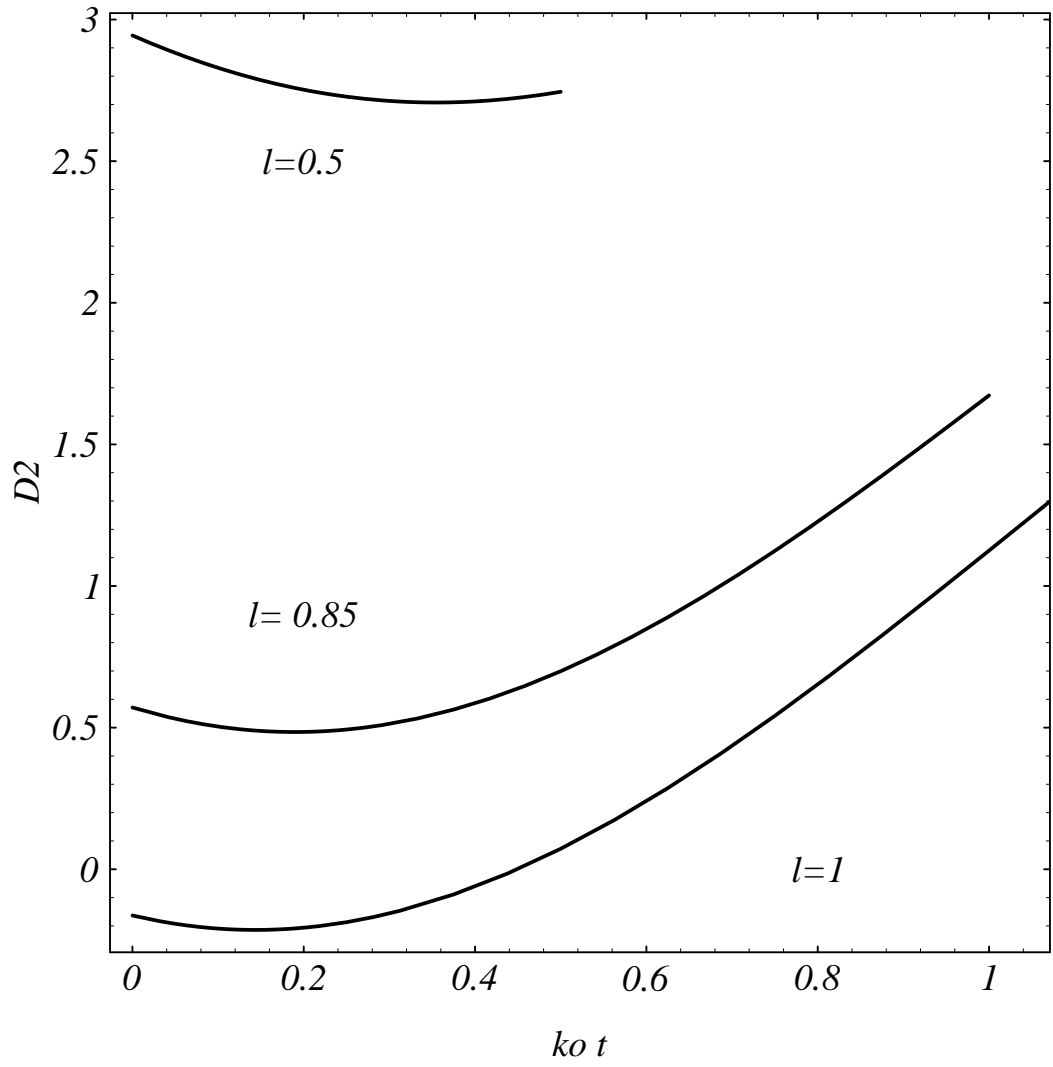
FIG.5. The same as Fig. 4 but with smaller values for $l$

\end{document}